\documentclass[12pt,a4,notitlepage,fleqn]{article}

\usepackage[verbose]{geometry}
\usepackage[symbol]{footmisc}
\usepackage[T1]{fontenc}
\usepackage{authblk}

\makeatletter

\newcommand{\fboxsubsec}[1]{
	\begin{flushleft}
		#1
	\end{flushleft}
	}
	\newcommand{\fboxsubsubsec}[1]{
	\begin{flushleft}
		#1
	\end{flushleft}
	}
\renewcommand{\subsection}{\@startsection{subsection}{2}{0pt}
	{1ex}
	{0.5ex}
	{\reset@font\it\fboxsubsec}
	}
\renewcommand{\subsubsection}{\@startsection{subsubsection}{2}{0pt}
	{1ex}
	{0.5ex}
	{\reset@font\fboxsubsubsec}
	}
\makeatother

\title{Discretion versus Policy Rules in Futures Markets:\\
A Case of the Osaka-Dojima Rice Exchange,\\
1914-1939}%

\author{Mikio Ito$^{a}$, \ Kiyotaka Maeda$^{b}$ \ and \ Akihiko Noda$^{c,d}$\thanks{\scriptsize Corresponding Author. E-mail: noda@cc.kyoto-su.ac.jp, Tel: +81-75-705-1510, Fax: +81-75-705-3227.}

{\scriptsize ${}^{a}$ \it Faculty of Economics, Keio University, 2-15-45 Mita, Minato-ku, Tokyo 108-8345, Japan}

{\scriptsize ${}^{b}$ \it Faculty of Economics, Seinan Gakuin University, 6-2-92 Nishijin, Sawara-ku Fukuoka 814-8511, Japan}

{\scriptsize ${}^{c}$ \it Faculty of Economics, Kyoto Sangyo University, Motoyama, Kamigamo, Kita-ku, Kyoto 603-8555, Japan}

{\scriptsize ${}^{d}$ \it Keio Economic Observatory, Keio University, 2-15-45 Mita, Minato-ku, Tokyo 108-8345, Japan}}
\date{This Version: \today}

\renewcommand\thefootnote{\arabic{footnote}}

\usepackage{graphicx} 

\usepackage[]{natbib}%
\usepackage{amsmath,amssymb}%
\usepackage{ascmac}%
\usepackage{multirow}%
\usepackage{lscape}%
\usepackage{subfigmat}

\usepackage{pifont}%
\usepackage{arydshln}%
\usepackage[format=hang]{caption}
\usepackage[all]{xy}
\usepackage{url}
\bibpunct{(}{)}{;}{a}{}{,}

\def\hsymbu#1{\smash{\lower1.7ex\hbox{\huge$#1$}}}

\newcommand{\citetapos}[1]{\citeauthor{#1}'s \citeyearpar{#1}}
\newcommand{\citeapos}[2]{\citeauthor{#1}'s (\citeyear{#2})}

\newcommand{\bs}{\texttt{\symbol{'134}}}
\def\cad#1{\texttt{\bs #1} & \csname #1\endcsname }

\begin{document}

\begin{titlepage}

\renewcommand{\thepage}{}
\renewcommand{\thefootnote}{\fnsymbol{footnote}}

\maketitle

\vspace{-10mm}

\noindent
 \hrulefill

\noindent
{\bfseries Abstract:} We investigate the relationship between market efficiency of rice futures transaction in Osaka and the Japanese government intervention in rice distributions by directly buying and selling rice during the interwar period, from the middle 1910s to 1939, considering the context of ``discretion versus rules.'' We use a time-varying VAR model to compare market efficiency and the government's actions over time. We found the two facts by featuring the time-varying nature of the market efficiency. First, the intervention with discretionary power disrupted the rice market and reduced market efficiency in the exchange. Second, the market efficiency improved in accordance with reduction in the government's discretionary power to operate the rice policy. When the government obtained the discretionary power to operate the policy regarding commodity market, the market efficiency often reduced. Conversely, even if the government implemented a large-scale intervention, the market efficiency improved when the government chose a systematic rule-like behavior following the law.\\

\noindent
{\bfseries Keywords:} Rice Futures Market; Market Efficiency; Time-Varying VAR Model; Rules; Descretion.\\

\noindent
{\bfseries JEL Classification Numbers:} N25; N45; G13; G18; G28; C22.

\noindent
\hrulefill

\end{titlepage}

\pagebreak

\section{Introduction}\label{osaka_sec1}
Economists have investigated how and to what extent each county's government intervened in the commodity market and industry. In particular, there is much literature on optimal monetary policies in a context of ``discretion versus rules'' for the last four decades. (See, for example, \citet{kydland1977rrt}, \citet{barrow1983rdr}, \citet{taylor1993dpr}, \citet{mccallum1993sam} and \citet{clarida1999smp}) The literature has demonstrated that policy rules are superior to discretion when the time inconsistency lesson by Kydland and Prescott is valid and when market participants believe the commitment to rules. As for their difference, in general, discretion is regarded as period-by-period optimization by the authority; a rule calls for period-by-period implementation of a contingent treatment that is selected from a generally applicable list for many decision periods. Following \citet{taylor1993dpr}, we understand that rule-like behavior implies systematic in the sense of ``methodical, according to a plan, and not casual or at random''. At the same time, we follow \citetapos{mccallum1993sam} idea that the authority ``must also design the systematic response pattern [so as] to take account of the private sector's expectational behavior'' at one time. Accordingly, this paper investigates the relationship between market efficiency of the Osaka-Dojima Rice Exchange and the Japanese government intervention in rice distributions by directly buying and selling rice from 1914 to 1939.

Being interested in a series of interventions into the rice exchanges in the prewar Tokyo and Osaka, we have already studied the relation between market efficiency and interventions in the rice markets using monthly data (see \citet{ito2016meg,ito2017fpe}). We paid our attention to the timing of interventions in the rice exchanges more than the nature of interventions. Nevertheless, considering the fact that the government's actions to the rice markets were different in nature at each period, we investigate the differences in this paper using daily data, only available for the Osaka-Dojima Rice Exchange, with more attentions. More specifically, we examine whether the rice prices in the exchange reflected them and its market efficiency would reduce or not when the government interventions in the exchange were systematic rule-like behavior and easy to be expected. However, the rice market in the prewar Japan is an unusual subject to study the government policies in the context of ``discretion versus rules'' since it was irrelevant to monetary policies. In particular, during the period that we address in this paper, the Japanese government struggled to find better policies for stable rice distributions and for controlling rice price; it attempted some actions: prohibiting traders with highly speculative transaction, directly buying and selling rice in the spot market, intervening in the rice exchange and establishing laws to facilitate the above actions. The changing policies of the prewar Japanese government provide us with lessons on which type of policy, discretionary or rule-like, in a commodity market had advantages for the objective. These policies had relation with a shift in the structure of world economy in interwar period.

Many governments of the countries that have their own colonies had strengthened their interventions from the First World War. During the WWI, the European countries raised their tariffs, and introduced the quantitative restrictions on trade and the system of import licenses. Despite the agreement that the international community removed the restriction in the Genoa Conference of 1922, any countries did not abolish their import prohibitions after the war (see \citet{kindleberger1989cpb}; \citet[pp.443--446]{findlay2007ppt}).In addition, they restricted foreign trade to protect the domestic industry from the Great Depression of 1929. For example, the US government enacted the Smoot-Hawley Tariff Act to perform the tariff protection in 1930 (see \citet{eichengreen2003ach}). The protection policies divided the world economy into blocks, and major countries depended on their colonial trading for supply of foods and raw materials during interwar period. These economic structure and policy in Japan were no exception in the same period.

Japan won the First Japanese-Sino War from 1894 to 1895, and Taiwan became the first Japanese colony. From the war, Japan expanded its territory by colonizing parts of the East Asia and Southeast Asia, and it became a unique Asian country which possessed colony. These colonies such as Taiwan and Korea supplied the food crops and industrial resources to the home islands from the turn of the twentieth century (see \citet[pp.82--85]{francks2015jed}). In particular, during interwar period, Japan enhanced a dependency on its colonial trading to a greater extent than European countries and formed its block economy in East Asia since the geographical distance between the metropole and the colony in Japan was closer than that of European countries (see \citet{okubo2007tbf} and \citet{hori2009hce}). In this block economy, rice was the main item which imported from the Japanese colonies to the home islands.

In Japan, rice had been a staple food since ancient times and had been a listed good in the commodity exchange since the early modern period; the Tokugawa Shogunate certified the {\it Dojima Kome Kaisho} (the Osaka-Dojima Rice Exchange) in 1730. From the same year to the 1930s before the Second World War in Japan, rice was a representative listed good in the futures market. In fact, there were 37 exchanges in 1930, and rice was traded in 25 exchanges including 19 exchanges which specialized in rice trading (see \citet[pp.4--5]{mci1931soe}). Accordingly, the rice policy had great significance for not only food supply policy but also commodity market policy. Especially, rice occupied an important position of price stabilization policy. Before the WWII, price fluctuation of rice strongly affected the change in general prices. Actually, 13 percent of the weight of the Tokyo Wholesale Price Index in the 1933 Base was the price of rice (see \citet[p.40]{boj1987hys}). Therefore, the Japanese government had implemented the rice policy and strengthened its rice policy after the WWI. During the WWI, Japan experienced industrialization and urbanization and sent its troops to Siberia. Under this situation, traders anticipated a rise in rice prices and cornered in the rice market. As a result, rice prices increased rapidly, and {\it Kome Sodo} (the rice riots) occurred nationwide in 1918. Reflecting on the disruption from the riot, the government strengthened its rice policy to adjust the balance of rice supply and demand. There were two characteristic points in the policy during interwar period in Japan.

First, the government compensated the rice shortage in the home islands of Japan with the imports of colonial rice. In 1920, the Governor-General of Korea implemented {\it Sanmai Zoshoku Keikaku} (the Plan to Increase in Rice Yield in Korea) to resolve the shortage (see \citet[pp.104--114]{yi2015eck}; \citet[p.5]{korea1922spi}). Second, the Japanese government directly bought and sold rice in the spot market according to the Rice Law in 1921 and the Rice Control Law in 1933. We have much literature for this point. Based on \citet{ohuchi1950faj} and \citet{suzuki1974rhc} which clarify the decision-making process of the rice policy, \citet{kawahigashi1990hrp} asserts that the rice policy reconciled conflicting interests between landowners and industrial factory owners. \citet{tama2013rmf} examines that the government purchase of rice encouraged fiercer competition between brands of rice from the 1920s. \citet{omameuda1993fpm} examines that the strengthening of rice policy caused to split the Ministry of Agriculture and Commerce into the Ministry of Agriculture and Forestry and the Ministry of Commerce and Industry in 1925. On the other hand, some literature focuses on the relationship between the rice policy and rice market exists. \citet{mochida1970drm} mentions that the rice policy which suppressed the price volatility of rice triggered declines in the rice futures market. At the same time, establishing the two laws above resulted in the implementation of government's policy clearer for rice traders than before. In particular, the Rice Control Law called for period-by-period implementation of a contingency formula regarding the rice market. However, the government intervention in market generally causes reduction of the market efficiency as much literature insists.

Considering the government's action to be added into an information set of market participants, one can apply a test of semi-strong efficiency in the line of \citet{fama1970ecm}. Since \citet{hansen1980fer} formalized a method for testing both weak and semi-strong forms of efficiency in foreign exchange markets, some studies in the context of a semi-strong form of efficiency have examined whether the rate of return is affected by some factors other than the lagged rates. For example, \citet{frenkel1980erp} examines the foreign exchange market efficiency under floating systems. By regressing the forecast error to other markets' forecast errors. \citet{goss1987wpp} studies whether Australian wool futures prices reflect available information including government actions. \citet{chance1985ssf} studies how the inflation announcement would affect the treasury bond futures market efficiency by using an event study and a dummy variable regression.

However, some literature indicates that the rice futures market was efficient during interwar period when the Japanese government often intervened. \citet{shizume2011fep} asserts that the rice futures market in Tokyo succeeded in providing a fine index of the expected price of physical rice from the late 1920s to the late 1930s. Using a generalized least squares (GLS) based time-varying VAR model, \citet{ito2016meg,ito2017fpe} investigates the time-varying market efficiency of the rice market from 1881 to 1932 and examines that the efficiency of the rice market improved in the 1920s. Conversely, \citet{ito2016meg,ito2017fpe} does not discuss the relationship between a change in the rice policy and the time-varying market efficiency of rice sufficiently since it mainly focuses on how an increase in rice imports from the Japanese colonies affected the price formation of rice in the home islands of Japan. Consequently, any literature does not investigate the time-varying efficiency of the rice market in the 1930s when the government highly strengthened its rice policy.

This paper studies how and to what extent the government interventions in the rice market affected the market efficiency of the Osaka-Dojima Rice Exchange over time periods, from 1914 to 1939. The Osaka-Dojima Rice Exchange was the largest of the exchanges which specialized in rice trading in Japan. In fact, its annual commission revenue from rice transactions was 900,766 yen while that of the second-largest rice exchange, the Nagoya Rice Exchange, was 165,366 yen in 1930 (see \citet[pp.4--5, 9--10]{mci1931soe}). Therefore, focusing on the Osaka-Dojima Rice Exchange, we precisely examine that the interrelation between the government’s directly buying and selling rice and a change in the efficiency of the rice market.

This paper is organized as follows. Section \ref{osaka_sec2} provides a brief historical review on rice policy the WWI in Japan. Section \ref{osaka_sec3} presents our method to study market efficiency varying over time based on, \citeapos{ito2014ism}{ito2014ism,ito2017aae} model. Section \ref{osaka_sec4} describes our daily dataset, covering the futures prices in the Osaka-Dojima Rice Exchange, and presents preliminary unit root test results. Section \ref{osaka_sec5} shows our empirical results using the time-varying VAR models and discuss the relation between the changing government's rice policy and time-varying market efficiency. Section \ref{osaka_sec6} concludes.

\section{Historical Review on Rice Policy in Interwar Japan}\label{osaka_sec2}

Since the 1890s, Japan had suffered from a shortage of rice because of rapidly increasing in rice demand. Especially, the Japan's shortage became more serious after the Russo-Japanese War from 1904 to 1905 along with its urbanization and industrialization (see \citet[pp.60--63, 80--83]{omameuda1993fpm}). Accordingly, the Japanese government implemented two countermeasures to suppress increasing in rice prices due to the shortage. First, the Ministry of Agriculture and Commerce forced the exchanges to accept imported rice as a deliverable in 1912. This order increased the supply of deliverable rice in the exchanges to decline futures prices (see \citet{ito2016meg}). Second, the Governor-General of Korea promoted japonica rice cropping in Korea. Imported rice including Korean rice had been indica variety while Japan's domestic rice was japonica variety; the former had differences in texture and taste from the latter. As a result, rice consumers in the home islands of Japan regarded Korean rice as an alternative grain. Under this situation, the second policy shrank the above quality differences between domestic and Korean rice (see \citet[pp.322--326]{tobata1939rek}). Nevertheless, the rice prices of all grades in Japan shot up in the late 1910s.

During the WWI, the Japanese government sent its troops to Siberia in 1918. This military action caused the speculative holding-off by rice traders who anticipated a rise in rice prices. The rice prices actually shot up, and the nationwide riots, {\it kome-sodo} (the rice riots), occurred in the same year. Accordingly, the Ministry of Agriculture and Commerce attempted to improve the policy regarding controls of both the price and the balance between supply and demand of rice. In 1920, the government consulted the {\it Rinji Zaisei Keizai Chosakai} (Extraordinary Economic and Finance Council) on the policy. In the next year, the Council recommended it to introduce a new system to adjust the balance of rice supply-and-demand, and the government enforced the {\it Beikoku Ho} (Rice Law) in April 1921 (see \citet[pp.100--145]{kawahigashi1990hrp}).

The Rice Law enabled the government to implement two instruments to control the balance between supply and demand of rice. Concretely, the government was allowed to trade rice in the spot market and alter import duty on foreign rice. However, the government could not buy and sell rice in order to stabilize rice price fluctuations since the Rice Law did not allow the government to intervene in the spot market to control rice prices directly. According to the Rice Law, when the government intended to intervene in the market, it had to estimate the supply and demand of rice in advance. Nevertheless, the law did not clearly state the criteria of government intervention. Then, while the government could hardly recognize how much it should buy or sell rice, rice traders could not expect when and to what extent it would intervene. As a result, the government could not conduct price-keeping operations with rice when Japan experienced rich harvest of rice. Accordingly, it amended the Rice Law to control the rice prices in March 1925 (see \citet[pp.198--202]{omameuda1993fpm}).

The amended Rice Law enabled the government to intervene in the spot market to control the rice prices. Figure 1 exhibits the futures prices of second nearest and differed contract transactions from 1914 to 1939.
\begin{center}
(Figure \ref{osaka_fig1} around here)
\end{center}
Figure \ref{osaka_fig1} indicates that the prices fell down from the middle to the late in the 1920s. In the same period, the Ministry of Agriculture and Forestry frequently bought the rice in the spot market to keep the prices . In short, the government intervened in the rice spot market in response to the price fluctuations of rice. Thus, the amendment of the Rice Law made it easier for the rice traders to forecast the intervention. However, it failed to keep the rice prices because it could not prevent the Korean rice imports from increasing. According to the amended Rice Law, the colonial rice including Korean rice was exempt from the control of import duty. Under the circumstances, whenever the government bought the rice in the spot market to keep the prices, rice traders in Korea, who anticipated an increase in rice prices by the government intervention, sent Korean rice to the home islands of Japan. As a result, the imports of Korean rice increased when the government conducted price-keeping operations, and the government failed to keep the prices (see \citet[pp.728--729]{hishimoto1938skr}). Accordingly, the Japanese government abolished the amended Rice Law and enforced the {\it Beikoku Tosei Ho} (Rice Control Law) in October 1933.

The government established the Rice Control Law to suppress the seasonal fluctuation of rice distribution. Concretely, the law required the government to set a price range by deciding maximum and minimum prices of rice in each year. If the physical rice prices deviated from the range, the government had to intervene in the spot market to adjust the prices (see \citet[pp.4--5]{maf1933eoe}). To put it differently, this law clearly stated the criteria of government intervention. The criteria under the amended Rice Law, which was stated clearer than that under the original Rice Law, did not denote the price range. Thus, the government was hardly able to declare when it began to buy or sell the physical rice in response to the price fluctuations of rice until September 1933. In contrast, even the traders easily expected when the government would intervene in the rice spot market after the establishment of the Rice Control Law. In the same period, the government attempted to strengthen its power to control the rice prices. Accordingly, it enforced the {\it Beikoku Jichi Kanri Ho} (Rice Self-Management Law) in September 1936 and administered the Rice Control Law and the Rice Self-Management Law from then on (see \citet[pp.3--17]{tsunoda1937ers}). Concretely, the Rice Self-Management Law required farmers and land owners to organize 25 {\it Beikoku Tosei Kumiais} (Associations for Management of Rice Distributions) in all corners of the home islands of Japan, Taiwan, and Korea, and these associations stored the rice whose quantity was decided by the government according to the Rice Self-Management Law. When the rice prices deviated from the price range which was the government set according to the Rice Control Law, the Ministry of Agriculture and Forestry could ordered the associations not to sell the stored rice. However, the government had never prevented the associations from selling the rice (see \citet[pp.845--849]{ota1938rpp}). Consequently, the Rice Self-Management Law did not affect the rice price formation.

From the same period, the Japanese government strengthened its wartime regime and economic control after it opened the war against China in July 1937. The rice trading was influenced by the circumstances. In April 1939, the {\it Beikoku Haikyu Tosei Ho} (Rice Distribution Control Law) was enforced. The sixth article of the law stated the government prohibited from the rice trading in futures. Accordingly, the government launched into preparations toward the abolishment of rice futures trading. In July of the same year, the government established the exclusive agent, Nihon Beikoku Co., Ltd., which would manage the rice market under the Rice Distribution Control Law. In the next month, the rice exchanges including the Osaka-Dojima Rice Exchange suspended their operations. Finally, Nihon Beikoku launched its operation and the rice traders could buy and sell only in the spot market which Nihon Beikoku managed from October 1939 (see \citet[pp.82--121]{itagaki1939erd}).

\section{The Model}\label{osaka_sec3}
This section provides a brief review of our method, which owes much of \citet{ito2014ism,ito2017aae}. Following their idea, we esimate  an vector autoregressive (VAR) model with time-varying coefficients. Then, we study time-varying nature  of market efficiency of Osaka-Dojima Rice Exchange using its daily data in relation to a series of rice law amendments. 

Suppose that ${p_t}$ is a price vector of rice futures with two different contract months at $t$ period. Our main focus is reduced to the condition 
\begin{equation}
E\left[ {x_{t}} \mid \mathcal{I}_{t-1}\right]=0,  \label{EMH}
\end{equation}%
where ${x_{t}}$ denotes a return vector of futures at $t$ period, that is, $i$-th componet of ${x_{t}}$ is $\ln {p_{i,t}}-\ln p_{i,t-1}$ for $i=1,2,3$.  In other words, all expected returns at $t$ period given the information set available at $t-1$ period are zero.

When $x_{t}$ is stationary, the Wold decomposition allows us to regard the time-series process of $x_{t}$ as 
\begin{equation*}
x_{t}=\mu + \Phi _{0}u_{t} + \Phi _{1}u_{t-1}+\Phi _{2}u_{t-2}+\cdots,
\end{equation*}%
where $\mu$ is the mean of $x_t$ and $\left\{ u_{t}\right\} $ follows an i.i.d. multivariate process with the mean of zero vector, and a covariance matrix of $\sigma^{2} I$, $\sum_{i=0}^{\infty}||\Phi_{i}' \Phi_{i}||<\infty$ with $\Phi_{0}=I$. Note that the efficient market hypothesis (EMH) holds if and only if $\Phi_i  = 0$ for all $i>0$. This suggests that how the market deviates from the efficient market reflects the impulse response, a series of $\{u_{t}\}$'s. In this paper, we shall construct an index based on the impulse response in order to investigate whether the EMH holds for the Osaka-Dojima Rice Exchange. 

The easiest way to obtain the impulse response is using an VAR model and algebraically computation of its coefficient estimates. Under some conditions, the vector return process $\{x_t\}$ of rice futures is invertible. We consider the following time-varying VAR($q$) model
\begin{equation}
x_{t}=\nu + A_{1}x_{t-1}+A_{2}x_{t-2}+\cdots+A_{q}x_{t-q}+\varepsilon_{t},  \label{arq}
\end{equation}
where $\nu$ is an intercept term and $\varepsilon_{t}$ is an multivariate error term with $E\left[\varepsilon_{t}\right]=0$, \ $E\left[\varepsilon_{t}^{2}\right]=\sigma_{\varepsilon}^{2} I $, and $E\left[\varepsilon_{t}\varepsilon_{t-m}\right]=0$ for all $m\neq 0$. We can use the idea of of \citet{ito2014ism,ito2017aae} when we measure market efficiency that would vary over time. Directly applying the Equation (7)  to our model, we obtain our degree computed through the VAR estimated coefficients,
$A_1,\cdots,A_q$, as follows. First, we compute a cumulative sum of the coeffient matrices of the impulse response,  
\begin{equation}\label{LongRunMult}
\Phi(1)=\left(I - A_1 - A_2 - \cdots - A_q\right)^{-1},
\end{equation}
Second, we define our degree of market efficiency,
\begin{equation}\label{LongRunMult}
\zeta=\sqrt{\mbox{max} \left[(\Phi(1) - I)'(\Phi(1) - I)\right]},
\end{equation} 
to measure the deviation from efficient market. Note that in the case of efficient market where $A_1=A_2=\cdots=A_q=0$, our degree $\zeta$ becomes zero; otherwise, $\zeta$ deviates from zero. That is why we call $\zeta$ the degree of market efficiency. When we find a large deviation of $\zeta$ from $0$ (both positive and negative), we can regard some deviation from one as an evidence of market inefficiency. Furthremore, we can construct this degree that would vary over time when we obtain time vrying estimates of the coefficients in Equation (\ref{arq}).

Adopting a method developed by \citet{ito2014ism,ito2017aae}, we estimate VAR coefficients at each period in order to obtain the degree defined in Equation (\ref{LongRunMult}) at each period. In practice, following their idea, we use a model in which all the VAR coefficients, except for the one that corresponds to the intercept term, $\nu$, follow independent random walk processes. That is, we suppose  
\begin{equation}
A_{l,t}=A _{l,t-1}+V_{l,t}, \ \ (l=1,2,\cdots,q), \label{rw_a}
\end{equation}
where an error term matrix $\left\{V_{l,t}\right\}$ ($l=1,2,\cdots,q$ and $t=1,2,\cdots,T$)  satisfies $E\left[V_{l,t}\right]=O$ for all $t$, \ $E\left[vec(V_{l,t})'vec(V_{l,t})\right]=\sigma_v^{2} I$ and $E\left[vec(V_{l,t})'vec(V_{l,t-m})\right]=O$ for all $l$ and $m\neq 0$. \citeapos{ito2014ism}{ito2014ism,ito2017aae} method allows us to estimate the time-varying VAR (TV-VAR) model:
\begin{equation}
x_{t}=\nu+A_{1,t}x_{t-1}+A_{2,t}x_{t-2}+\cdots+A_{q,t}x_{t-q}+\varepsilon_{t},  \label{tv_arq}
\end{equation}
together with Equation (\ref{rw_a}). 

In order to conduct statistical inference on our time-varying degree of market efficiency, we apply a residual bootstrap technique to the TV-VAR model above. In practice, we build a set of bootstrap  samples of the TV-VAR estimates under the hypothesis that all the TV-VAR coefficients are zero. This procedure provides us with a (simulated) distribution of the estimated TV-VAR coefficients assuming the rice futures return processes are generated under the efficient market hypothesis. The, we can compute the corresponding distributions of the impulse response and degree of market efficiency. Finally, by using confidence bands derived from such simulated distributions, we conduct statistical inference on our estimates and detect periods when the Osaka-Dojima rice exchange experienced market inefficiency.

\section{Data}\label{osaka_sec4}

In this paper, we conduct a dataset of daily rice futures prices in the Osaka-Dojima Rice Exchange from September 1, 1914 through August 25, 1939. For the rice futures data in the Osaka-Dojima Rice Exchange, we utilize weighted average daily values for two contract months is available: a second nearest contract (two months), and a deferred contract (three months).\footnote{We except a series of nearby contract (one month) to conduct our dataset. Because there exists many missing values in a nearby contract.} The dataset consists of a primary source in {\it Monthly Reports on Osaka-Dojima Rice Exchange}.\footnote{We find the primary source of daily futures prices in the Osaka-Dojima Rice Exchange at Kansai University Library (\url{http://opac.lib.kansai-u.ac.jp/}) which is located in Osaka, Japan.} Our dataset ignores sunday, different national holidays and regular exchange closings. For missing values due to irregular holidays by imperial events and temporary exchange closings because of natural catastrophe or casualty, we fill in with values obtained by spline interpolation. And we take the first difference of the natural log of rice futures prices to compute the ex-post returns. 

For our estimations, all variables that appear in the moment conditions should be stationary. To confirm whether the variables satisfy the stationarity condition, we apply the \citetapos{elliott1996eta} augmented Dickey-Fuller generalized least squares (ADF-GLS) test. We employ the modified Bayesian information criterion (MBIC) instead of the modified Akaike information criterion (MAIC) to select the optimal lag length. This is because, from the estimated coefficient of the detrended series, $\hat\psi$, we do not find the possibility of size-distortions (see \citet{elliott1996eta}; \citet{ng2001lls}). Table \ref{osaka_table1} shows the results of the unit root test with descriptive statistics for the data. The ADF-GLS test suggests that all variables do not contain a unit root at the 1\% significance level.
\begin{center}
(Table \ref{osaka_table1} around here)
\end{center}

\section{Empirical Results}\label{osaka_sec5}
\subsection{Preliminary Results}

We first assume a time-invariant VAR($q$) model with constant and employ \citetapos{schwarz1978edm} Bayesian information criteria (SBIC) to select the optimal lag order in our preliminary estimations. Table \ref{osaka_table2} summarizes our preliminary results for a time-invariant VAR($q$) model using the whole sample.
\begin{center}
(Table \ref{osaka_table2} around here)
\end{center}

Table \ref{osaka_table2} also shows that the cumulative sum of the VAR estimates for the case of differed contract month is smaller than the second nearest case. It shows that the differed contract month market is the most efficient. This preliminary result coincides with \citetapos{ito2017fpe} time-invariant estimation, which concludes that the deferred contract month market is the most efficient. 

 We investigate whether the parameters are constant in the above VAR($q$) models using \citetapos{hansen1992a} parameter constancy test under the random parameters hypothesis. Table \ref{osaka_table2} also presents the test statistics; we reject the null of constant parameters against the parameter variation as a random walk at the 1\% significance level. Therefore, we estimate the time-varying parameters of the above VAR models to investigate whether gradual changes occur in interwar Osaka-Dojima Rice Exchange. These results suggest that the time-invariant VAR($q$) model does not apply to our data and that the TV-VAR($q$) model is a better fit.

From historical view point, the Japanese government strengthened its rice related policy during the interwar period as we mentioned in Section \ref{osaka_sec2}. The policy potentially affected the price formation of rice since it enabled the government to intervene in the rice market. Accordingly, we estimate the degree of time-varying market efficiency using the TV-VAR model in the next subsection.

\subsection{Time-Varying Market Efficiency and Historical Interpretation}
\begin{center}
(Figure \ref{osaka_fig2} around here)
\end{center}

Figure \ref{osaka_fig2} shows that the degree of time-varying market efficiency gradually improved over time. Until the early 1920s, rice transaction was conducted in the almost inefficient market until the 1910s. However, the market efficiency improved from the middle 1920s. Specifically, the transaction was conducted in efficient market in the following periods: from December 1925 to May 1928, from December 1929 to September 1930, from October 1930 to October 1931, from May to November in 1932, from February to June in 1933, and after November 1933. In brief, the market became nearly efficient after the amendment of the Rice Law in 1925. Finally, it was efficient after the establishment of the Rice Control Law in 1933. In fact, according to the standard deviations for the four periods: 0.170, 0.099, 0.060 and 0.049, we find that the time-varying degree became less and less volatile over time. Accordingly, we discuss why the efficiency improved in accordance with the amendment of the Rice Law and the establishment of the Rice Control Law paying attention to discretion versus rules in the interwar Japanese rice policy. These laws were aimed at adjusting the rice demand-and-supply balance and controlling rice price as mentioned in Section \ref{osaka_sec2}. Then, we investigate trends in the rice futures prices and the rice supply in interwar Japan.

According to Figure \ref{osaka_fig1}, the futures prices fluctuated wildly even under the Rice Law and the Rice Control Law. This fluctuation depended largely on the change in the rice supply.
\begin{center}
(Figure \ref{osaka_fig3} around here)
\end{center}

Figure \ref{osaka_fig3} presents Japan's annual supply volume of rice per capita from 1914 to 1939. The statistics, which were calculated by the Ministry of Agriculture and Forestry, consisted of initial stock, production, and foreign trade including colonial trade. The figure indicates that the trend of rice supply also fluctuated greatly, and we find that this fluctuation had relation with the variation in the degree of time-varying market efficiency under the amended Rice Law. According to Figures \ref{osaka_fig1} and \ref{osaka_fig3}, from 1925 to 1933, whenever the ratio of the supply volume to its five years moving average volume attained above or under three percent, the efficiency tended to reduce. However, before and after this period, the efficiency was irrelevant to the rice supply. Before the amendment of the Rice Law, the transaction was conducted in the almost inefficient market even when the supply volume did not deviate from its five years moving average volume. On the other hand, after the establishment of the Rice Control Law, the futures market was always efficient even when the supply volume varied wildly.

In summary, the trend of the efficiency changed when the government altered the rice policy. Under this policy, the Japanese government attempted to adjust the rice demand-and-supply balance and control the rice price by frequently intervening in the rice market. Accordingly, we argue the relationships between the inefficient of rice market and the government intervention from the next subsection.

\subsubsection{Confusion of the Rice Exchange System in the 1910s}
In the 1910s, the futures transaction was conducted in the almost inefficient market. During the same period, the Japanese government forced the rice exchanges to accept the delivery of the imported rice. As \citet{ito2016meg} mention in detail, the government interventions involving the delivery of imported rice often reduced the efficiency of the rice futures markets.

In prewar Japan, the sellers in the rice exchanges frequently delivered physical rice to clear their futures transactions. When they delivered physical rice, the exchanges basically permitted them to use only domestic rice. In contrast, the delivery rule was changed when the rice supply became insufficient since the industrialization and urbanization progressed in the 1910s. In the same period, the rice price increased rapidly as shown in Figure \ref{osaka_fig1}, and the government experimented with suppressing the rice price to change the delivery rule which permitted the sellers to deliver the imported rice such as Taiwanese and Korean rice in the exchanges. Consequently, buyers in the exchanges significantly faced uncertainty about the rice delivery since they were liable to obtain the imported rice whose quality was different from the domestic rice. This situation caused a disruption in the rice exchanges, and the futures price failed to be a fine index of the expected price of rice in the spot market as \citet{ito2016meg} assert. In fact, Figure \ref{osaka_fig2} shows that the transaction was conducted in the almost inefficient market until 1919. Specifically, the market efficiency reduced rapidly from the late 1917 to the early 1919. This circumstance resulted from the following three factors.

First, the Ministry of Agriculture and Commerce enforced the {\it Bori Torishimari Rei} (the Ministerial Ordinance on Anti Excessive Profit) to suppress the rice price on September 1 1917. Under this ministerial ordinance, if a trader cornered or held off selling the commodity to obtain excessive profits, the Minister of Agriculture and Commerce could prohibit him from trading of the commodity. In fact, the government punished 18 rice traders including nine traders in the Osaka-Dojima Rice Exchange for violation of the ministerial ordinance from October 1917 to August 1918 (see \citet[pp.266--269]{ota1938rpp}). However, the government did not clearly state the criteria of conflict with the ministerial ordinance, and rice traders could not expect how the government would regulate them. As a result, the rice traders in the exchanges feared the punishment by the government, and refrained from the dealing. In fact, during period spanning June through November, the total trading volume in the Osaka-Dojima Rice Exchange decreased from 55 million {\it koku} in 1916 to 50 million {\it koku} in 1917 (see \citet[p.21]{dojima1916sar} and \citet[p.25]{dojima1917sar}).\footnote{{\it Koku} is a unit of rice trading volume in Japan. One {\it koku} is equal to 180.39 liters.} The Osaka-Dojima Rice Exchange mentioned this situation, ``After the government punished the rice buyers in the Osaka-Dojima Rice Exchange according to the Ministerial Ordinance on Anti Excessive Profit, the traders were thrown into some confusion. Consequently, the buyers refrained from the dealing, and the rice futures price fell down.''\footnote{See \citet{dojima1918mr} for detail.} That is, when the rice price increased, the buyers in the exchange could not trade freely since they were scared of the punishment according to the ministerial ordinance. This situation caused the exchange not to collect sufficient information on rice demand, and the market efficiency reduced.

Second factor was the suspension of the futures trading in the exchange. In 1918, the Ministry of Agriculture and Commerce condemned the rice futures as the gambling to raise the rice price. Then, the government forbade the rice futures trading in Osaka in April and from July to August.
\begin{center}
(Figure \ref{osaka_fig4} around here)
\end{center}

Figure \ref{osaka_fig4} shows that the days of the suspending transactions in the Osaka-Dojima Rice Exchange from 1914 to 1939. According to Figure \ref{osaka_fig4}, the futures trading in the exchange was suspended for many days in 1918. This suspension of the trading caused the situation that rice traders could not exchange information on supply and demand of rice sufficiently. 

Third, the Ministry of Agriculture and Commerce inspected the Osaka-Dojima Rice Exchange and checked all the traders' account books in June 1918. This inspection also suppressed the rice futures trading since the buyers were scared of the punishment by the government (see \citet[p.11]{dojima1918sarb}).

These three factors interrupted the exchange to collect the information on supply and demand of rice and reduced the efficiency of the futures market (see Figure \ref{osaka_fig2}). However, the rice price decreased in the early 1919 (see Figure \ref{osaka_fig1}), and the government eased the regulation of the rice futures trading. In fact, the government did not punish traders for violation of the ministerial ordinance from September 1918, and the days of the suspending transactions decreased drastically in 1919 (see Figure \ref{osaka_fig4}). As a result, the market efficiency improved temporarily in the middle 1919.

\subsubsection{Rice Futures under the Original Rice Law}

From the late 1920 to the early 1922, the time-varying market efficiency reduced. In September 1920, the Ministry of Agriculture and Commerce felt the necessity to adjust rice supply and demand fundamentally, and began to design the Rice Law in secret (\citet[p.73]{hasumi1967byh}). In the next month, {\it Yomiuri Shimbun} (Yomiuri Newspaper) reported that the government planned to establish the Rice Law.\footnote{See \citet{yomiuri1920na} for detail.} In response to this situation, the traders suspected that the government attempted to prevent the rice price from declining at the same period when the supply volume of rice increased and the rice price declined (see Figures \ref{osaka_fig1} and \ref{osaka_fig3}).  In fact, the semi-annual report of the Osaka-Dojima Rice Exchange mentioned, ``Although the rice supply excessed, the rice price increased as the government has prepared the Rice Law. The traders suspected that the government intervened in the rice spot market (see \citet[p.7]{dojima1921sara}).'' The traders' expectation about the government intervention interrupted the process of price formation; the information on the supply and demand of rice did not directly reflected in the futures price. As a result, the market efficiency reduced from the late 1920.

In April 1921, the Rice Law was established. The establishment of the Rice Law did not improve the market efficiency immediately. According to Figure \ref{osaka_fig2}, the futures transaction was conducted in the inefficient market until 1925. This situation resulted from the fact that traders guarded against the government intervention under the Rice Law. Kai Irie, the bureaucrat of the Ministry of Agriculture and Commerce, stated that the traders paid close attention to the government's trading in the rice spot market (\citet[p.71]{rice1961ami}). In April 1921, Yomiuri Shimbun mentioned, ``The traders expect that the government will buy the rice in early summer, and the futures price of rice increases.''\footnote{See \citet{yomiuri1921mt} for detail.} That is, the traders attempted to expect when the government intervened in the rice market. Nevertheless, they could not expect accurately when the government would buy or sell the rice in the spot market. The Rice Law allowed the government to buy and sell in the rice spot market to adjust the balance of rice supply and demand. However, it did not clearly state the criteria of government intervention. In fact, Yasushi Hasumi, a head of Agricultural Policy Division, Agricultural Bureau, Ministry of Agriculture and Forestry from 1927 to 1929, mentioned in 1933.

\begin{quote}
``The government had often traded in the rice spot market of adjust the rice demand-and-supply balance. However there were no criteria of the intervention. Consequently, it was difficult for us to declare that when we began to trade rice in response to price fluctuation of rice.''\footnote{See \citet{ota1938rpp} for detail.}
\end{quote}

Accordingly, anyone could not forecast the government interventions. Under this situation, whenever the government intervened in the rice market, the efficiency reduced.
\begin{center}
(Figure \ref{osaka_fig5} around here)
\end{center}

Figure \ref{osaka_fig5} shows that the volume of the government's buying and selling of rice in the spot market from 1914 to 1939. According to this figure, the government intervened in the rice market frequently after the establishment of the Rice Law. Under the original Rice Law from 1920 to 1925, the government bought the rice in bulk in June 1921 and the beginning of 1923 when the rice price declined. On the other hand, it continuously sold the rice from the middle of 1923 to the middle of 1925 when the rice price increased (see Figure \ref{osaka_fig1}). As a result, according to Figure \ref{osaka_fig2}, the efficiency reduced at the same period and improved in 1922 when the government did not buy and sell the rice in bulk. However, the futures transaction was still conducted in the inefficient market. As we mentioned in Section \ref{osaka_sec2}, the government intervened in the spot market to control the balance between supply and demand of rice under the original Rice Law; it could not intervene in the market to control the rice price. Accordingly, the government always intervened in the market only after estimating the supply and demand of rice (see \citet[pp.198--201]{omameuda1993fpm}). Under the circumstances, the government could not buy and sell the rice flexibly in response to the change in the supply and demand of rice, and rice traders could not forecast the intervention. Consequently, if the government did not intervene in the market, the transaction was conducted in the inefficient market under the original Rice Law because the market could not form the price reflecting the future interventions.

\subsubsection{Rice Futures under the amended Rice Law}

\begin{figure}[bp]
 \caption{Futures Prices of Second Nearest and Differed Contract Transactions}\label{osaka_fig1}
 \centering
 \includegraphics[scale=0.8]{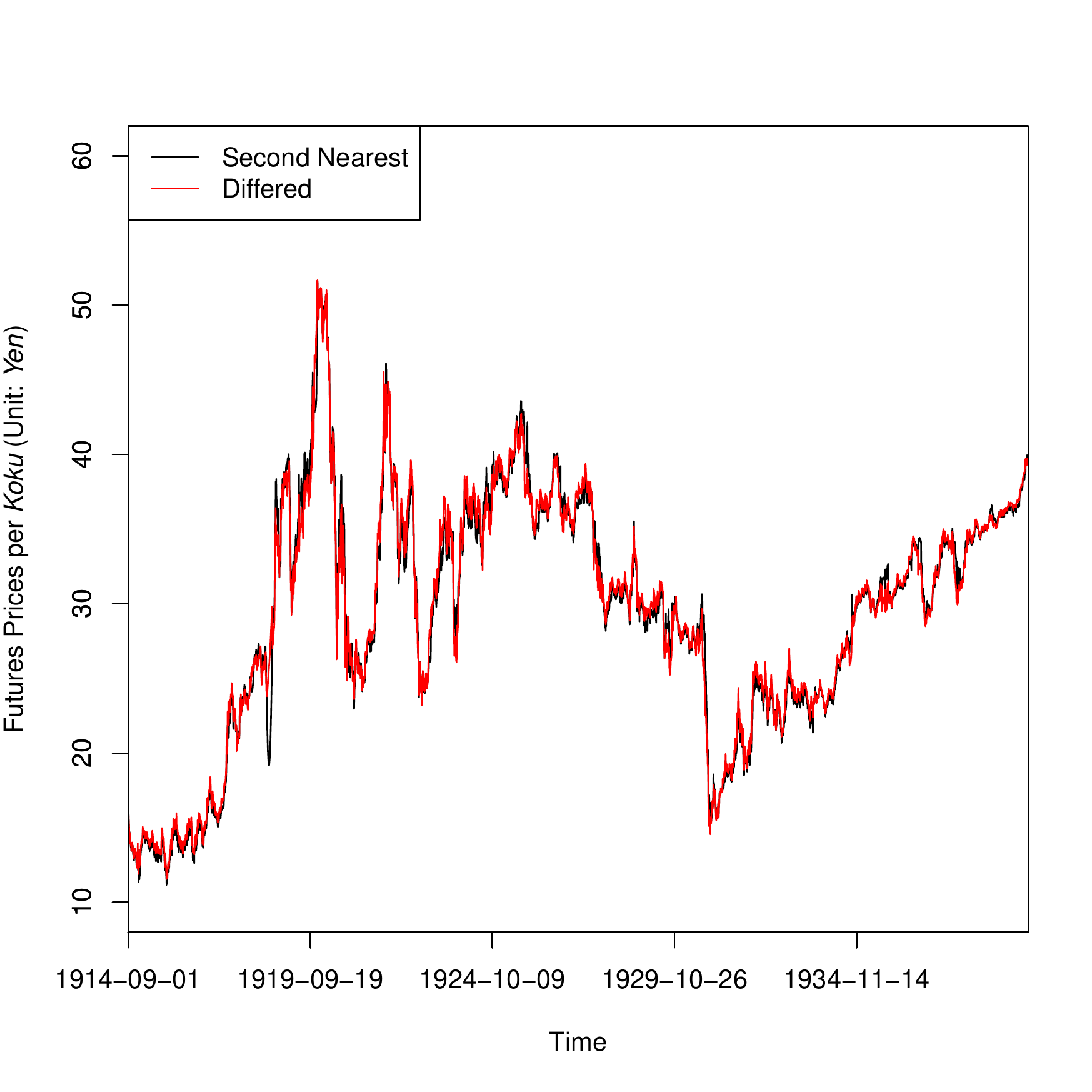}
\vspace*{3pt}
{
\begin{minipage}{370pt}
 \footnotesize
 \begin{itemize}
  \item[\underline{Data Sources}:] \citet{dojima1914mr,dojima1915mr,dojima1916mr,dojima1917mr,dojima1918mr,dojima1919mr,dojima1920mr,dojima1921mr,dojima1922mr,dojima1923mr,dojima1924mr,dojima1925mr,dojima1926mr,dojima1927mr,dojima1928mr,dojima1929mr,dojima1930mr,dojima1931mr,dojima1932mr,dojima1933mr,dojima1934mr,dojima1935mr,dojima1936mr,dojima1937mr,dojima1938mr,dojima1939mr}
  \item[\underline{Note}:] The shaded areas are inefficient markets in our sample period.
 \end{itemize}
 \end{minipage}}%
\end{figure}

\begin{figure}[bp]
 \caption{The Time-Varying Degree of Market Efficiency}\label{osaka_fig2}
 \centering
 \includegraphics[scale=0.75]{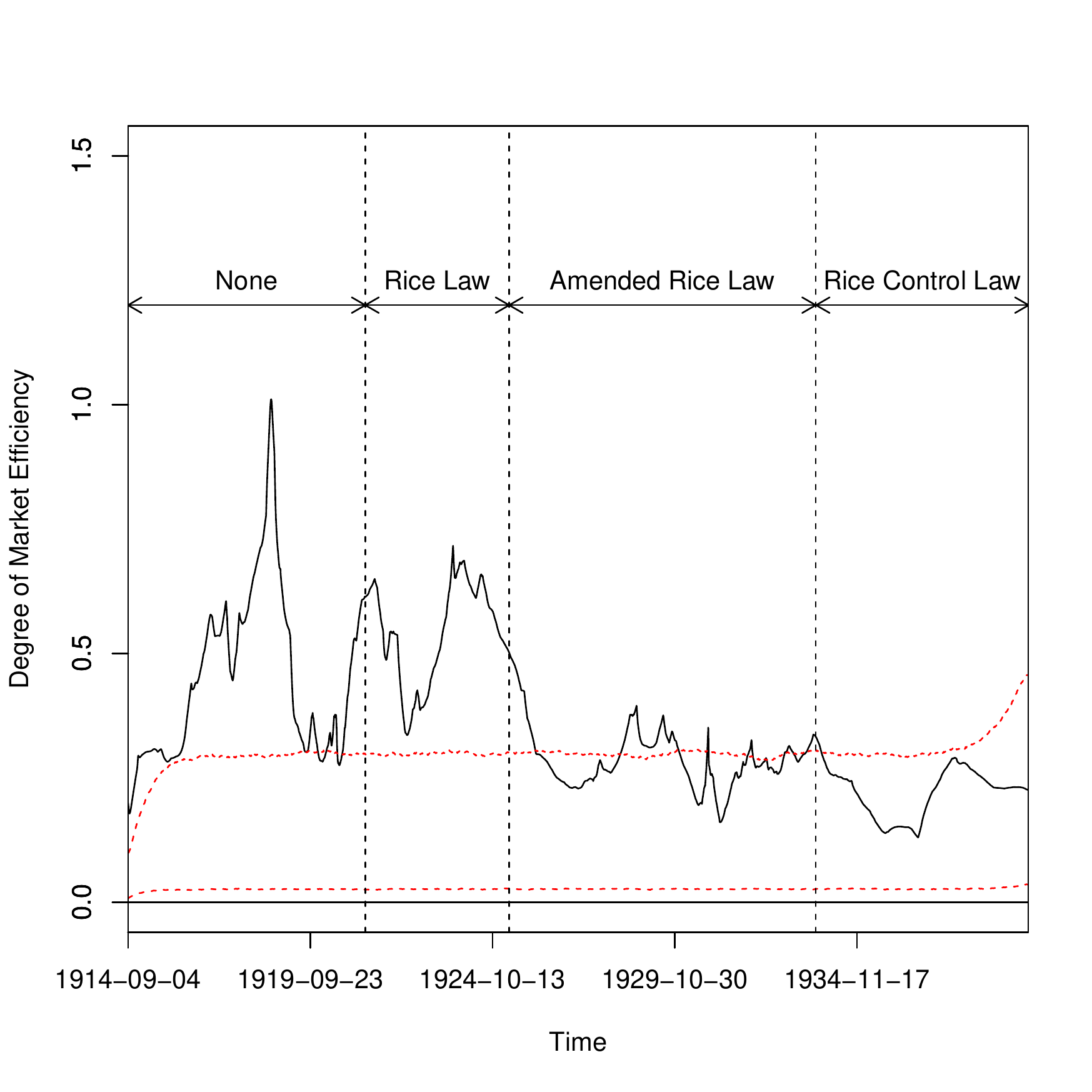}
  \vspace*{3pt}
{\small
\begin{minipage}{450pt}
{\underline{Notes:}}
\begin{itemize}
\item[(1)] The dashed red lines represent the 95\% confidence bands of the time-varying spectral norm in the case of an efficient market.
\item[(2)] We ran bootstrap sampling 5000 times to calculate the confidence bands.
\item[(3)] R version 3.4.3 was used to compute the estimates.
\end{itemize}
\end{minipage}}
\end{figure}

As we mentioned in Section \ref{osaka_sec2}, the government amended the Rice Law in March 1925. After the amendment of the law, the government intervened in the spot market to control the rice price. This change made it easier for rice traders to forecast the intervention since the government could buy and sell the rice flexibly in response to the price fluctuation of rice. In short, when the rice price varied greatly, the traders could expect how the government would intervene. In other words, when the price was stabilized under the amended Rice Law, the traders did not need to pay attention to the intervention, and the futures market could collect information on the supply and demand of rice. In fact, from 1926 to the following year when the rice price exhibited relatively less volatility than the early and the late 1920s, the government did not frequently buy and sell the rice (see Figures \ref{osaka_fig1} and \ref{osaka_fig5}). During the same period, according to Figure \ref{osaka_fig2}, the market efficiency improved. In contrast to the situation under the original Rice Law, the futures transaction was conducted in the efficient market when the government did not intervene in the rice market under the amended Rice Law. However, the efficiency began to reduce in the late 1927.

\begin{figure}[bp]
 \caption{Rice Supply in Japan (1914-1939)}\label{osaka_fig3}
 \centering
 \includegraphics[scale=0.8]{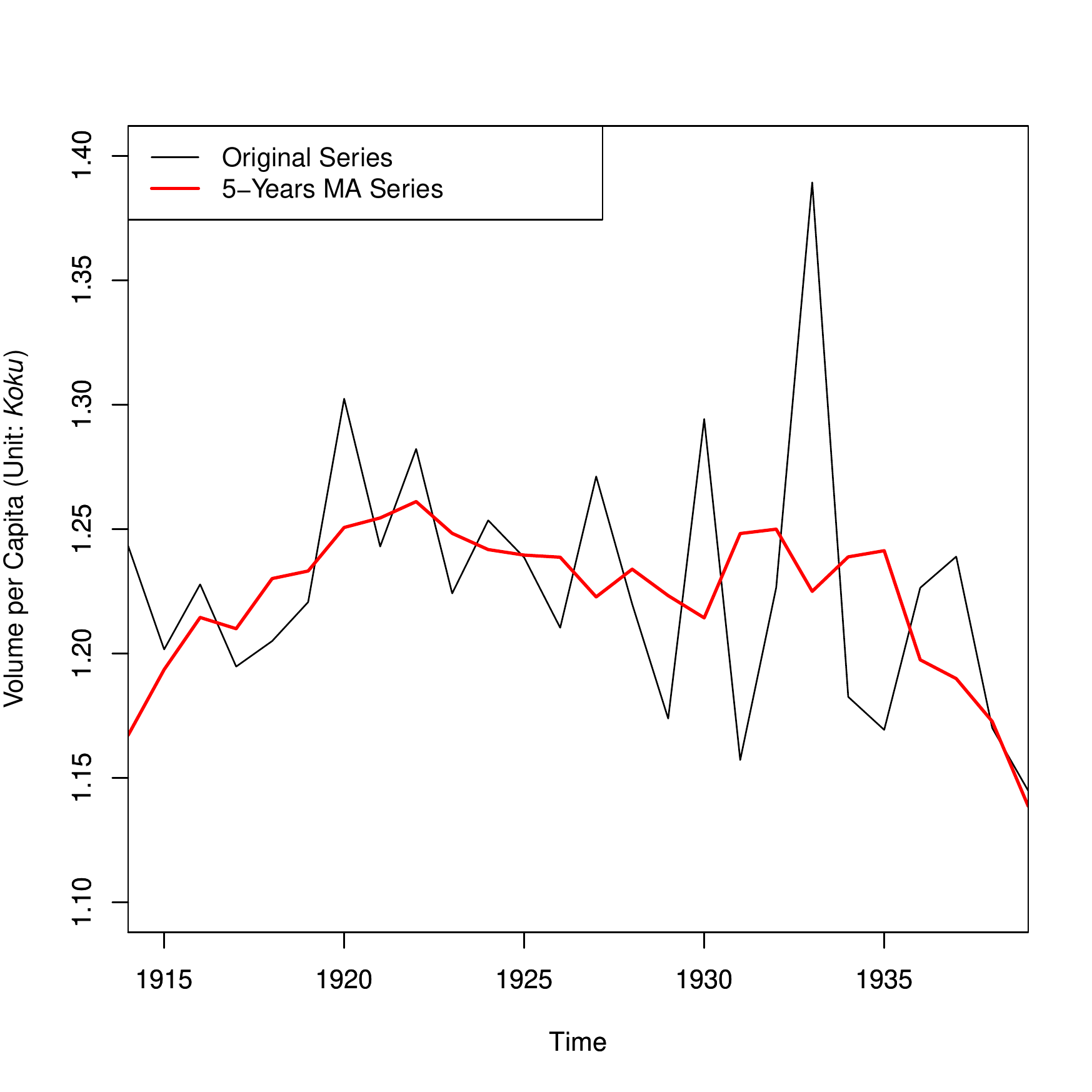}
\vspace*{3pt}
{
\begin{minipage}{370pt}
 \footnotesize
 \begin{itemize}
  \item[\underline{Data Sources}:] \citet[pp.52--53]{maf1944df}.
 \end{itemize}
\end{minipage}}%
\end{figure}

\begin{figure}[bp]
 \caption{Days of Suspending Transactions (1914-1939)}\label{osaka_fig4}
 \centering
 \includegraphics[scale=0.8]{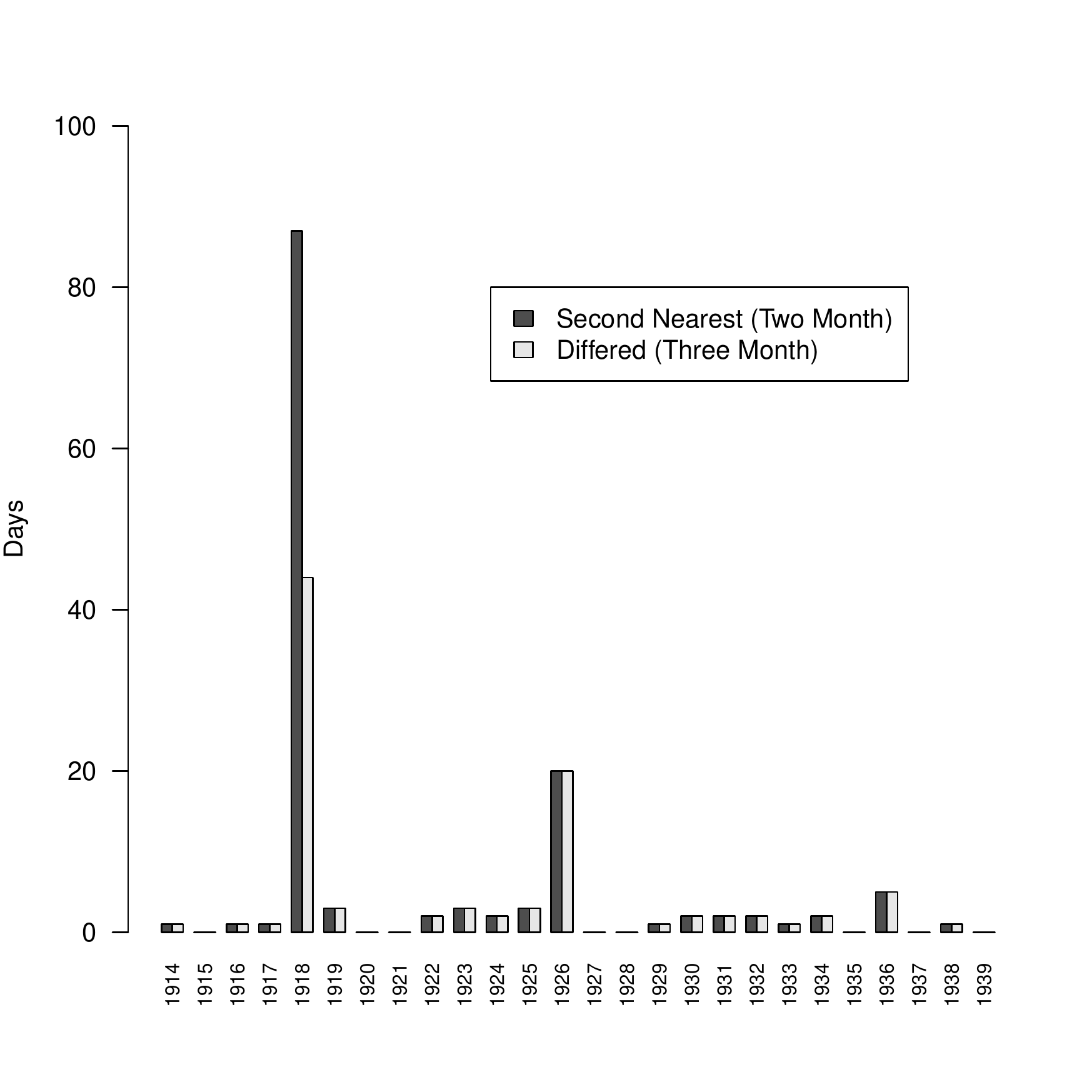}
 \vspace*{3pt}
{
\begin{minipage}{370pt}
 \footnotesize
 \begin{itemize}
  \item[\underline{Data Sources}:] \citet{dojima1914mr,dojima1915mr,dojima1916mr,dojima1917mr,dojima1918mr,dojima1919mr,dojima1920mr,dojima1921mr,dojima1922mr,dojima1923mr,dojima1924mr,dojima1925mr,dojima1926mr,dojima1927mr,dojima1928mr,dojima1929mr,dojima1930mr,dojima1931mr,dojima1932mr,dojima1933mr,dojima1934mr,dojima1935mr,dojima1936mr,dojima1937mr,dojima1938mr,dojima1939mr}
 \end{itemize}
\end{minipage}}
\end{figure}

\begin{figure}[bp]
 \caption{Volume of Purchase and Sales of Rice by the Government in the Spot Market}\label{osaka_fig5}
 \centering
 \includegraphics[scale=0.55]{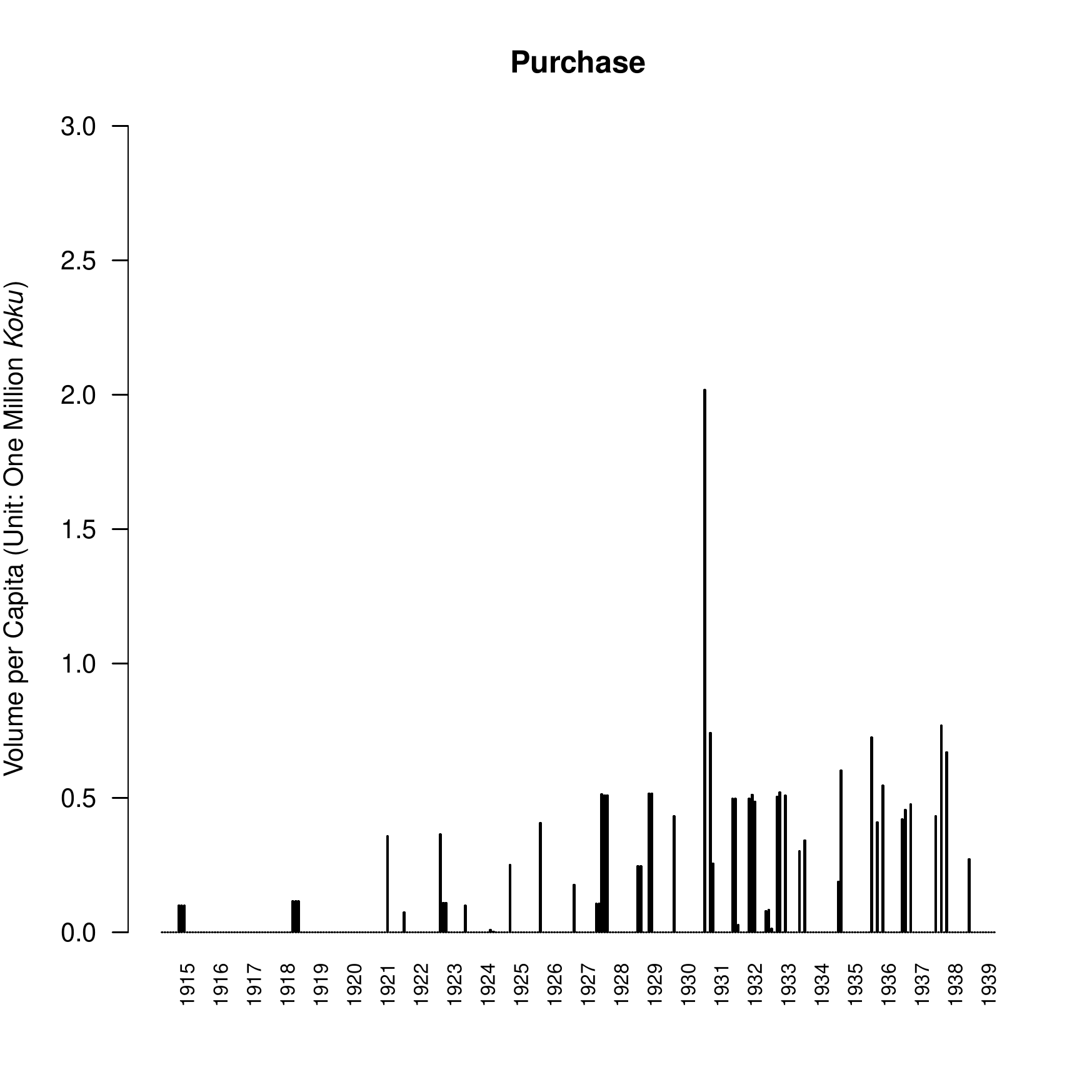}
 \includegraphics[scale=0.55]{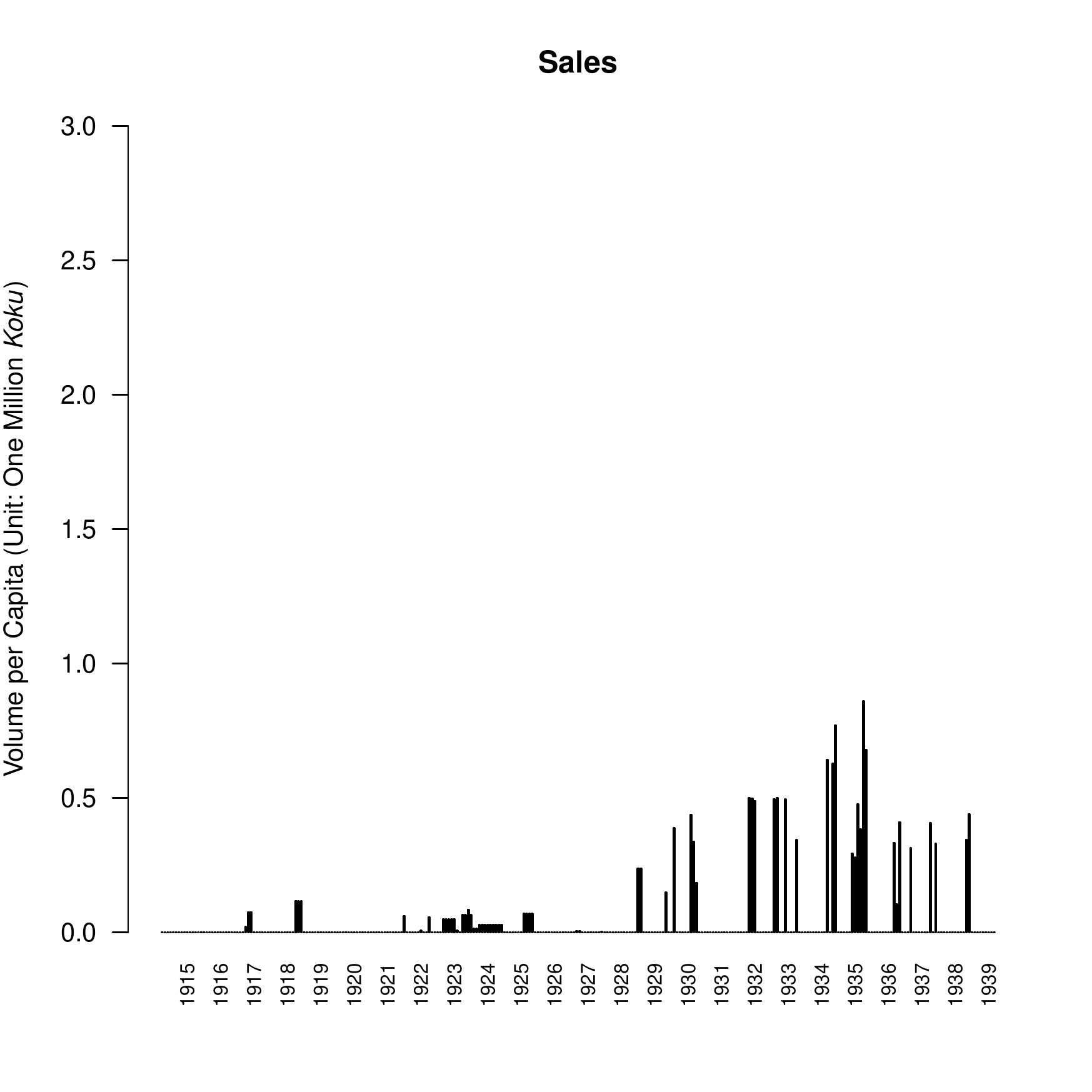}
  \vspace*{3pt}
{
\begin{minipage}{370pt}
\footnotesize
 \begin{itemize}
  \item[\underline{Data Sources}:] \citet[pp.275--318]{maf1944df}.
 \end{itemize}
\end{minipage}}%
\end{figure}

From 1927, the supply volume of rice deviated from its five years moving average volume, and the futures price fluctuated wildly. In 1927, the volume of rice supply was the largest after the introduction of the Rice Law (see Figures \ref{osaka_fig1} and \ref{osaka_fig3}). Accordingly, the government bought a large amount of rice to prevent rice prices from falling in the late 1927, and the efficiency began to reduce. However, this intervention failed to prevent rice prices from declining since the volume of rice imported from Korea increased in 1927 (see Figures \ref{osaka_fig1}, \ref{osaka_fig2}, and \ref{osaka_fig5}).

The Japanese government issued {\it Beikoku Shoken} (Rice Bill) to raise a fund for purchase of rice under the Rice Law, and it amounted to 56,683,460 yen of the Rice Bill in 1927. In the same year, the annual average price of physical rice was 35.23 yen per {\it koku} (see \citet[p.88]{maf1932hrl}). Accordingly, the government could buy 1,608,954 {\it koku} of rice, and the volume of the government's buying reached a ceiling in 1927 (see Figure \ref{osaka_fig5}). This ceiling on the buying was 2.6\% of the rice production in the year. If the volume of rice production decreased by 2.6\%, percentage change of the production volume from the average year would be 1.1\% in 1927 (see \citet[pp.52--53]{maf1944df}). This supposition suggests that the government was able to avoid falling rice price in 1927. However, the rice farmers in Korea also experienced abundant crops, and the rice import from Korea increased rapidly in the same year. In fact, its volume amounted to 7,068,709 {\it koku} while the corresponding past five years average of Korean rice imports was 4,709,055 {\it koku} from 1922 to 1926 (see \citet[pp.52--53]{maf1944df}). As the rice import from Korea increased by 50.1\%, the domestic rice supply also increased in 1927 (see Figure \ref{osaka_fig3}). Under this situation, the government failed to avoid falling rice price; the rice price in Japan declined rapidly in 1927. Accordingly, the Japanese government considered to constrain the rice imports and exports in Korea, and the rice traders attempted to expect when the government enforced the Rice Law partially in Korea. This expectation had a potential impact on price formation of rice in Osaka.

In the 1920s, a large portion of rice cropped in Korea was exported to the home islands of Japan; rice consumers in Korea basically bought foreign rice since the Korean rice was more expensive than the rice imported from foreign countries. On the other hand, rice supply in Osaka depended heavily on the Korean rice imports. In 1927, the monthly average volume of Korean rice stocks was 237,120 {\it koku} while that of domestic rice stocks was 123,706 {\it koku} in Osaka (see \citet{dojima1927ar}). Under this situation, the Japanese government implemented restrictions on rice trading in Korea to avoid falling rice price in the home islands of Japan (see \citet[pp.234--251]{omameuda1993fpm}). Concretely, the government enforced the Rice Law partially in Korea in February 1928, and the Governor-General of Korea operated the import license system of foreign rice from March to August in 1928 (see \citet[pp.396--397]{ota1938rpp} and \citet[p.88]{hasumi1957hmf}). In the same period, rice traders expected that these measures in Korea caused the decrease in the amount of Korean rice and the increase in rice price in Osaka. In fact, the Osaka-Dojima Rice Exchange reported in 1928, ``From January, most traders in the exchange expected that the government would enforce the Rice Law in Korea. As a result, the trade in futures became active, and the rice price increases.''\footnote{See \citet[p.7]{dojima1928mr} for detail.} This situation made it difficult for the rice traders to predict the rice supply from Korea. Furthermore, the traders could not predict the supply of domestic rice in the same year.

In 1928, the Osaka-Dojima Rice Exchange reported that the futures price of rice increased since the stored rice in the government's warehouses suffered from insect damage (see \citet{dojima1928ar}). In the end of the previous year when the rice price declined, the government bought the rice as we mentioned above. Consequently, the government's inventory quantity of rice increased by 56\% from 1927 to 1928. However, the government could not store the large quantity of rice properly (see \citet[pp.656--657]{ota1938rpp}). This failure decreased the supply volume of domestic rice.

In short, the rice traders could not predict the supply volume of Korean and domestic rice because of the government's action and failure. As a result, although the supply volume of rice did not deviate from its five years moving average in 1928, the futures transaction was conducted in the inefficient market (see Figures \ref{osaka_fig2} and \ref{osaka_fig3}).

In contrast to the rice supply in 1928, the supply volume of rice deviated from its five years moving average from 1929 under the amended Rice Law, and the rice price varied greatly (see Figures \ref{osaka_fig1} and \ref{osaka_fig3}). Accordingly, the government frequently intervened in the rice spot market. In this period, whenever the government bought or sold rice in bulk, the efficiency reduced (see Figures \ref{osaka_fig2} and \ref{osaka_fig5}). For example, Japan recorded bumper crop of rice in 1930, and it led to a dramatic increase in the supply volume of rice. As a result, the rice price declined drastically in the same year (see Figures \ref{osaka_fig1} and \ref{osaka_fig3}). In response to these circumstances, the government bought the rice in bulk in the spot market to avoid falling rice price in the late 1930 (see Figure \ref{osaka_fig5}). Furthermore, the rice price had been suppressed until the late 1931 because of the record harvest in 1930, although Japan experienced poor harvest of rice in the next year. Accordingly, the government intervened in the rice spot market again in the late 1931. These interventions reduced the efficiency as the case in the late 1927. Nevertheless, the rice price did not increase substantially until 1933 since the supply volume of rice increased from 1931 to 1933 (see Figures \ref{osaka_fig1} and \ref{osaka_fig3}). In the same period, according to Figure \ref{osaka_fig5}, the government frequently bought the rice in the spot market when the rice price declined and sold it when the rice price increased. These frequent interventions also reduced the efficiency from 1932 to the following year (see Figure \ref{osaka_fig2}). Therefore, the government interventions under the amended Rice Law heavily affected the price formation in the Osaka-Dojima Rice Exchange.

\subsubsection{Rice Futures under the Rice Control Law}

The Rice Control Law was established in November 1933. This law aimed at controlling the rice price and the seasonal fluctuation of rice distribution (see \citet[pp.728--729]{ota1938rpp}). The government presented the bill of the Rice Control Law to the Diet in February 1933. Actually, the government attempted to control the seasonal fluctuation of rice distribution before the presenting the bill. In October 1932, the government issued the Imperial Ordinance No.297. This ordinance allowed the government to buy, sell, process, and store the Korean and Taiwanese rice to control distribution volume of imported rice. In fact, the government bought and sold the Korean rice in bulk after the issuance of the Imperial Ordinance. Especially, the government intensively intervened from July to October in 1933 and sold 275,897 {\it koku} of Korean rice during the same period (see \citet[pp.71--72]{maf1942lrr}). This intervention caused the decrease in the Korean rice price. The four months average price of Korean rice decreased by 8.7\% from March–June to July–October in 1933 while that of domestic rice decreased by 3.4\% during the same period (see \citet[pp.98--99]{boj1987hys}). That is, the price trend of Korean rice did not fit with that of domestic rice. Consequently, the price trend of the deliverable commodity was different from that of the trading commodity in the Osaka-Dojima Rice Exchange. This difference disrupted the futures transaction, and the efficiency reduced in 1933.

\begin{figure}[bp]
 \caption{Monthly Net Volume of Rice by the Government}\label{osaka_fig6}
 \centering
 \includegraphics[scale=0.55]{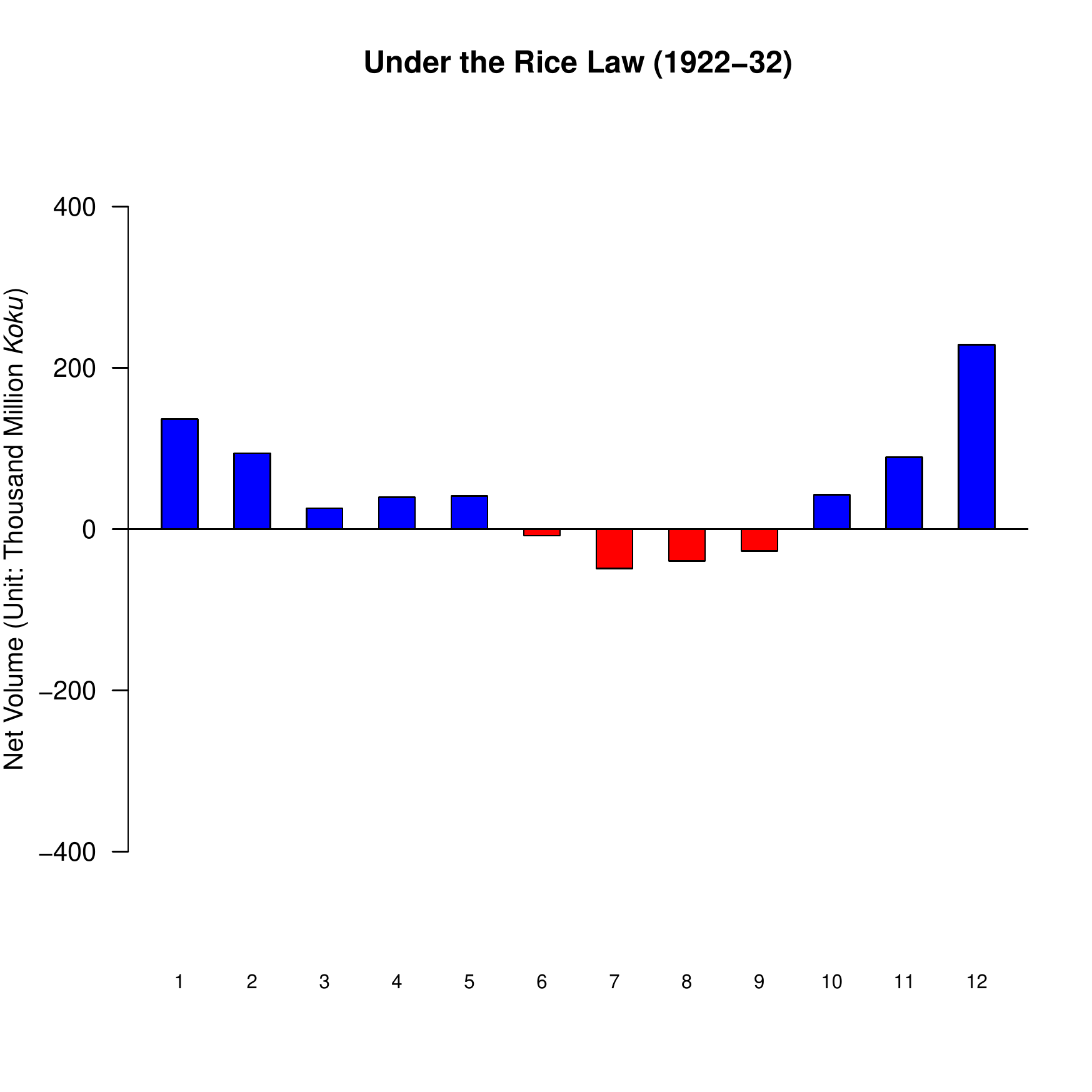}
 \includegraphics[scale=0.55]{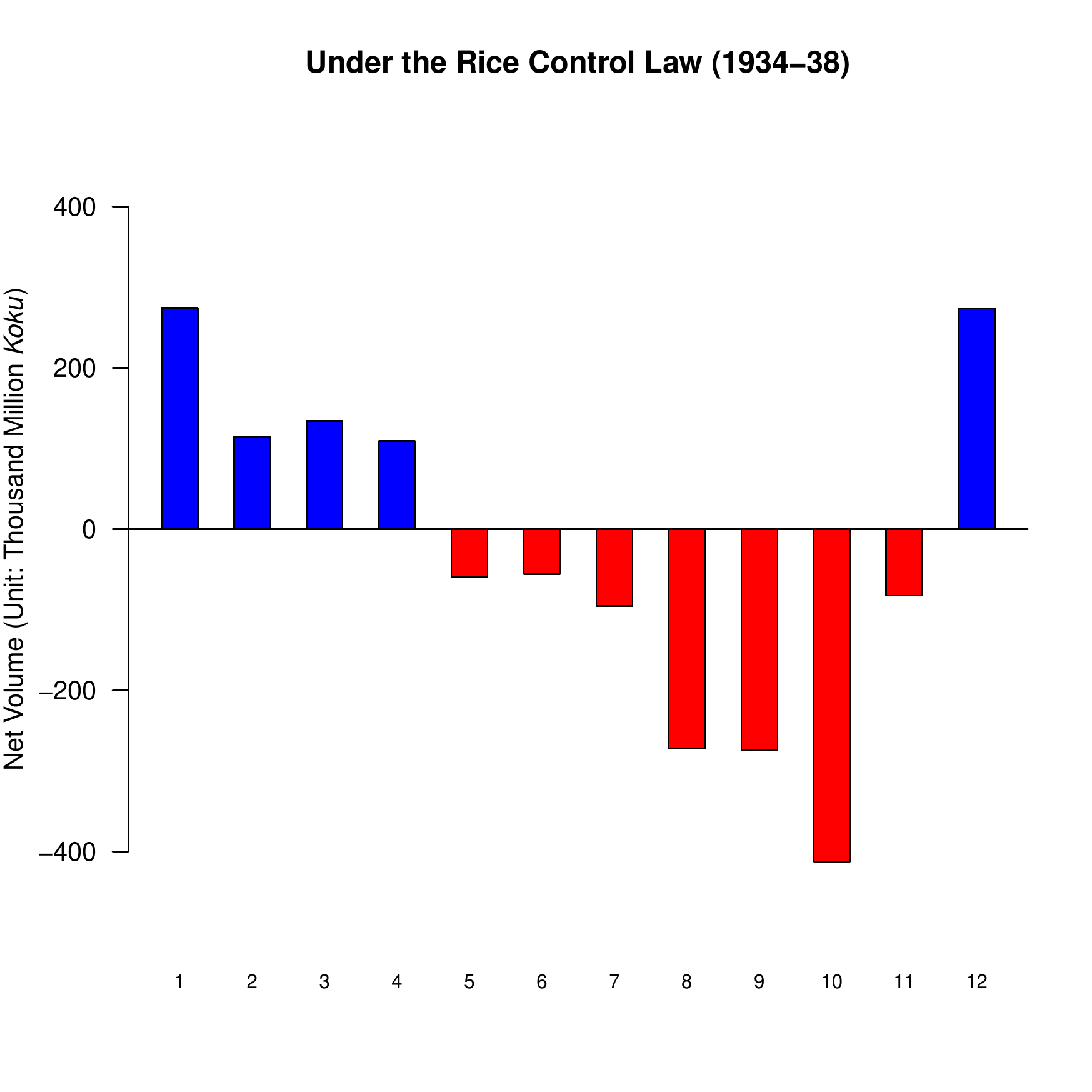}
  \vspace*{3pt}
{
\begin{minipage}{370pt}
 \footnotesize
 \begin{itemize}
  \item[\underline{Data Sources}:] \citet[pp.281--318]{maf1944df}.
  \item[\underline{Note}:] Monthly net volume of rice is calculated by deducting the monthly average volume of sales from that of purchase.
 \end{itemize}
\end{minipage}}
\end{figure}

From the establishment of the Rice Control Law in November 1933, the market continued to be efficient (see Figure \ref{osaka_fig2}). This situation resulted from the fact that the Rice Control Law clearly stated the criteria of government intervention. The law required the government to set a price range by deciding maximum and minimum prices of rice in each year (see \citet[p.731]{ota1938rpp}). If the price in the rice spot market deviated from the range, the government had to sell or buy rice to control the rice price. As a result, the government began to adjust the volume of buying and selling seasonally after the establishment of the Rice Control Law. Figure \ref{osaka_fig6} shows that the monthly average net volumes of government purchase under the Rice Law and the Rice Control Law.
\begin{center}
(Figure \ref{osaka_fig6} around here)
\end{center}

According to Figure \ref{osaka_fig6}, under the Rice Control Law, the government bought the rice in a harvest season when the rice price declined and sold it in a between-crop season when the rice price increased. Consequently, the traders could expect when the government would intervene in the rice spot market. As a result, under the Rice Control Law, the price formation in futures reflected the information on the government interventions. The discretionary of the government intervention reduced, and the efficiency in the futures market improved.

\section{Conclusion}\label{osaka_sec6}
Considering policies' nature in the context of ``discretion versus rules,'' we argue how and to what extent the Japanese government intervened in the rice futures market in Osaka during the interwar period when many countries' governments strengthened their interventions in markets. Featuring the Japanese government's constant concern about stable distributions of rice from the late 1910s to 1933, we found the two facts. First, the intervention with discretionary power disrupted the rice market and reduced market efficiency in \citetapos{fama1970ecm} sense in the exchange Second, the market efficiency improved in accordance with reduction in the government's discretionary power to operate the rice policy.

In 1917, the Ministry of Agriculture and Commerce enforced the Ministerial Ordinance on Anti Excessive Profit to suppress rice prices. However, because the government did not clearly state the criteria of conflict with the ministerial ordinance, rice traders could not expect when and how the government would punish them. Thus, the ordinance hampered rice traders' free trade when rice price increased. At this period, the government did not behave according to any rule and had some discretionary power. As a result, the Osaka-Dojima Rice Exchange failed to collect sufficient information on rice demand to help traders hedge their price risks of rice. This situation continued in the early 1920s.

In 1921, the Japanese government established the Rice Law to intervene in the rice spot markets. From the same year, the Ministry of Agriculture and Commerce directly bought and sold rice in the spot market to adjust the balance of rice supply and demand. The law did not cleary state the criteria of government intervention, and the government could intervene in the market on a discretionary basis. However, the law did not allow the government to trade to control the rice price, and the Ministry of Agriculture and Commerce always intervened only after estimating the supply and demand of rice. Consequently, the government intervention on a discretionary basis was not flexible, and rice traders could not forecast the intervention. Under the circumstance, the rice market failed to form the price which factored the information on future government interventions, and the efficiency of the rice futures market in the Osaka-Dojima Rice Exchange often reduced. Conversely, the amendment of the Rice Law in 1925 and the establishment of the Rice Control Law in 1933 restricted the government's discretionary power to operate the policy. These changes in rice policy altered the government's behavior in the rice market, and the market efficiency gradually improved.

The amended Rice Law allowed the government to intervene to control the rice price. Accordingly, the Ministry of Agriculture and Forestry bought and sold the rice flexibly in response to the price fluctuation of rice. In short, the amendment of the Rice Law made the timing of the government intervention clearer than that under the original Rice Law. After the amendment of the Rice Law, rice traders did not need to pay attention to the intervention when the rice price was stabilized, and the futures market could collect information on the supply and demand of rice. As a result, the futures transaction was conducted in the efficient market when the government did not intervene in the rice market. The restriction on the government's discretionary power significantly improved the market efficiency. However, the market efficiency reduced when the government intervened in the rice market under the amended the Rice Law since the law did not clearly state the criteria of government intervention. The traders was hardly able to forecast when the government began to buy or sell the rice in response to the price fluctuation of rice. In contrast to this situation, the market continued to be efficient from the establishment of the Rice Control Law in 1933.

Under the Rice Control Law, the government had to set a price range of rice in each year. If the price diviated from the range, the law required the government to intervene in the rice market. Accordingly, the Ministry of Agriculture and Forestry bought rice in a harvest season when the rice price declined and sold it in a between-crop season when its price increased. That is, the Rice Control Raw diminished the government's discretionary power to operate the rice policy because rice traders could expect interventions. Consequently, the rice market could form the price reflecting the future government interventions.

In brief, when the government obtained the discretionary power in the sense of \citet{taylor1993dpr} to operate the policy regarding commodity market, the market efficiency often reduced. Conversely, even if the government implemented a large intervention, the market efficiency improved when the government chose a systematic rule-like behavior following the law in the sense of \citet{mccallum1993sam}. The actions to the market performed a vital role in determining the market efficiency. The strengthening of policy on the market did not always reduced the market efficiency during interwar period in Japan.

\section*{Acknowledgments}

We would like to thank Shigehiko Ioku, Kozo Kiyota, Kris Mitchener, Chiaki Moriguchi, Tetsuji Okazaki, Minoru Omameuda, Masato Shizume, Yasuo Takatsuki, Yoshiro Tsutsui, and Tatsuma Wada for their helpful comments and suggestions. We would also like to thank seminar and conference participants at Keio University; the Japanese Economics Association 2017 Autumn Meeting; the 86th Annual Conference of the Socio-Economic History Society; and the 92nd Annual Conference of the Western Economic Association International for helpful discussions. We also thank the Japan Society for the Promotion of Science for their financial assistance, as provided through the Grant in Aid for Scientific Research Nos.17K03809 (Mikio Ito), 17K03863 (Kiyotaka Maeda), and 15K03542 (Akihiko Noda). All data and programs used for this paper are available on request.


\setcounter{table}{0}
\renewcommand{\thetable}{\arabic{table}}
 
\clearpage

\begin{table}[tbp]
\caption{Descriptive Statistics and Unit Root Tests}\label{osaka_table1}
\centering
\begin{tabular}{clcccccccccccc}\hline\hline
 &  &  & Mean & SD & Max & Min &  & ADF-GLS & Lags & $\phi$ &  & $\mathcal{N}$ & \\\cline{4-7}\cline{9-11}\cline{13-13}
 & $R_{sn,t}$ &  & 0.0001  & 0.0119  & 0.1072  & $-0.1425$  &  & $-13.8452$  & 14  & 0.2166  &  & 7411 &  \\
 & $R_{d,t}$ &  & 0.0001  & 0.0116  & 0.1048  & $-0.0907$  &  & $-12.0752$  & 15  & 0.2266  &  & 7411 &  \\\hline\hline
\end{tabular}
\vspace*{5pt}
{\small
\begin{minipage}{425pt}
{\underline{Notes:}}
\begin{itemize}
\item[(1)] ``ADF-GLS'' denotes the ADF-GLS test statistics, ``Lags'' denotes the lag order selected by the MBIC, and ``$\hat\phi$'' denotes the coefficients vector in the GLS detrended series (see equation (6) in \citet{ng2001lls}).
\item[(2)] In computing the ADF-GLS test, a model with a time trend and a constant is assumed. The critical value at the 1\% significance level for the ADF-GLS test is ``$-3.42$''.
\item[(3)] ``$\mathcal{N}$'' denotes the number of observations.
\item[(4)] R version 3.4.3 was used to compute the statistics.
\end{itemize}
\end{minipage}}%
\end{table}

\clearpage

\begin{table}[tbp]
\caption{Time-Invariant VAR Estimations}\label{osaka_table2}
\begin{center}\resizebox{8cm}{!}{
\begin{tabular}{lllccc}\hline\hline
 &  &  & $R_t^{sn}$ & $R_t^{d}$ & \\\cline{4-5}
 & \multirow{2}*{Constant} &  & 0.0001 & 0.0001 & \\
 &  &  & [0.0002] & [0.0001] &  \\
 & \multirow{2}*{$R_{t-1}^{sn}$} &  & 0.0072 & 0.1343 & \\
 &  &  & [0.0581] & [0.0495] & \\
 & \multirow{2}*{$R_{t-1}^{d}$} &  & 0.1740 & 0.0188 & \\
 &  &  & [0.0526] & [0.0537] & \\\hline
 & ${\bar R}^2$ &  & 0.0307 & 0.0235 & \\
 & $L_C$ &  & \multicolumn{2}{c}{342.4114} &  \\\hline\hline
\end{tabular}}\\
\vspace*{5pt}
{\resizebox{10cm}{!}{
\begin{minipage}{350pt}
{\underline{Notes:}}
 \begin{itemize}
\item[(1)] ``${\bar{R}}^2$'' denotes the adjusted $R^2$, and ``$L_C$'' denotes \citetapos{hansen1992a} joint $L$ statistic with variance.
\item[(2)] \citetapos{newey1987sps} robust standard errors are in brackets.
\item[(3)] R version 3.4.3 was used to compute the estimates and the test statistics.
 \end{itemize}
\end{minipage}}}%
\end{center}
\end{table}

\clearpage

\clearpage

\clearpage

\clearpage

\clearpage

\clearpage

\end{document}